\documentclass[fleqn,usenatbib]{mnras}

\usepackage[T1]{fontenc}
\usepackage{ae,aecompl}
\usepackage{hyperref}
\usepackage{adjustbox}

\usepackage{graphicx}	
\usepackage{amsmath}	
\usepackage{amssymb}	

\title{An Investigation of Four Chemically Peculiar Stars with Photometric Periods below 12 Hours}

\author[S. H{\"u}mmerich et al.]{
Stefan H{\"u}mmerich,$^{1,2}$\thanks{E-mail: ernham@rz-online.de}
Klaus Bernhard,$^{1,2}$
Ernst Paunzen,$^{3}$
Franz-Josef Hambsch,$^{1,2,4}$
\newauthor
Terry Bohlsen,$^{1,5,6}$
Jonathan Powles$^{5,7}$
\\
$^{1}$American Association of Variable Star Observers (AAVSO), Cambridge, USA\\
$^{2}$Bundesdeutsche Arbeitsgemeinschaft f{\"u}r Ver{\"a}nderliche Sterne e.V. (BAV), Berlin, Germany\\
$^{3}$Department of Theoretical Physics and Astrophysics, Masaryk University, Kotl\'a\v{r}sk\'a 2, 611 37 Brno, Czech Republic\\
$^{4}$Vereniging Voor Sterrenkunde (VVS), Brugge, Belgium\\
$^{5}$Southern Astro Spectroscopy Email Ring (SASER)\\
$^{6}$Mirranook Observatory, Boorolong Rd, Armidale, NSW, 2350, Australia\\
$^{7}$40 Hensman Street, Latham, ACT, 2615, Australia\\
}

\date{Accepted XXX. Received YYY; in original form ZZZ}

\pubyear{2016}

\begin{document}
\label{firstpage}
\pagerange{\pageref{firstpage}--\pageref{lastpage}}
\maketitle

\begin{abstract}
We present an investigation of three chemically peculiar (CP) stars and one CP star candidate which exhibit photometric periods below 12 hours. New spectroscopic observations have been acquired which confirm the peculiar nature of all objects. HD 77013 and HD 81076 are classical CP1 (Am) stars; HD 67983 is a marginal CP1 (Am:) star, and HD 98000 is a CP2 (Ap) star. We have procured observations from the ASAS-3 and SuperWASP archives and obtained additional photometry in order to verify the results from the sky survey data. We have derived astrophysical parameters and investigated the positions of our target stars in the $M_{\rm Bol}$ versus $\log T_\mathrm{eff}$ diagram, from which information on evolutionary status has been derived. We present period analyses and discuss each object in detail. From the available data, we propose pulsational variability as the underlying mechanism for the variability in HD 67983, HD 77013 and HD 81076, which offer the opportunity to study the interaction of atomic diffusion and pulsation. HD 67983 and HD 77013 exhibit multiperiodic variability in the $\gamma$ Doradus frequency realm; HD 81076 is a $\delta$ Scuti star. The CP2 star HD 98000 exhibits monoperiodic variability with a frequency of f\,$\approx$\,2.148 c/d (P\,$\approx$\,0.466 d), which we interpret as the rotational period. If this assumption is correct, HD 98000 is the $\alpha^{2}$ Canum Venaticorum (ACV) variable with the shortest period hitherto observed and thus a very interesting object that might help to investigate the influence of rotational mixing on chemical peculiarities.
\end{abstract}




\section{Introduction} \label{introduction}

About 15\% of upper main sequence stars between spectral types early B to early F are characterised by atmospheric chemical abundances that differ significantly from the solar pattern. Accordingly, these objects are referred to as chemically peculiar (CP) stars. The observed peculiarity is believed to be produced by selective processes (radiative levitation, gravitational settling) operating in calm radiative atmospheres \citep{2000ApJ...529..338R}. CP stars are not confined to a particular evolutionary stage \citep{2006A&A...450..763K} and are natural atomic and magnetic laboratories. Furthermore, because of their unusual composition characterised by significant under- or overabundances of several elements, they are ideally suited as testing grounds for the evaluation of model atmospheres \citep{2009A&A...499..567K}.

CP stars are intrinsically slow rotators, and it is thought that the meridional currents induced by the rapid rotation of hot main sequence stars will counteract the effects of the segregation of chemical elements, thus effectively preventing the formation of the characteristic abundance anomalies of the CP stars \citep[e.g.][]{2010A&A...511L...7M}. Therefore, an upper limit for rotational velocities is expected, beyond which no chemical peculiarities due to the above mentioned interplay of radiative diffusion and gravitational settling are expected. Establishing this upper limit is an important task, which might contribute to the understanding of the influence of rotational mixing on chemical peculiarities.

According to \citet{1974ARA&A..12..257P}, CP stars are commonly subdivided into four classes: metallic line (or Am) stars (CP1), magnetic Ap stars (CP2), HgMn stars (CP3), and He-weak stars (CP4). The CP1 stars are A- and early F-type stars and are defined by the discrepancies found in the spectral types derived from the strength of the Ca II K line, the hydrogen lines, and the metallic lines. In comparison to the spectral types derived from the hydrogen lines, the Ca II K-line types appear too early and the metallic-line types too late, i.e. Sp(K)\,$\leq$\,Sp(H)\,$\leq$\,Sp(m) -- with, obviously, the requirement Sp(K)\,<\,Sp(m) \citep{1940ApJ....92..256T}. In classical CP1 stars, Sp(K) and Sp(m) differ by five or more spectral subclasses; stars in which the difference is less are often referred to as 'marginal Am' (or Am:) stars.

CP1 stars do not show strong, global magnetic fields \citep{2010A&A...523A..40A} and are characterised by underabundances of calcium and scandium and overabundances of the iron-peak and heavier elements. CP1 stars are mostly members of binary systems with orbital periods between 2\,$\leq$\,$P\textsubscript{orb}$\,$\leq$\,10 d, and their rotational velocities are believed to have been reduced by tidal interactions, which has enabled diffusion to act \citep{2009AJ....138...28A}.

The CP2 stars are distinguished by their strong, globally organised magnetic fields that range up from about ~300 Gauss \citep{2007A&A...475.1053A} to several tens of kiloGauss \citep{2011IAUS..273..249K}. Below the above mentioned threshold, global magnetic fields are thought to be sheared by differential rotation. CP2 stars are notorious for exhibiting a nonuniform distribution of chemical elements, which is likely connected to the presence of the magnetic field and results in the formation of spots and patches of enhanced element abundance \citep{1981A&A...103..244M}. These inhomogeneities are responsible for the strictly periodic changes observed in the spectra and brightness of many CP2 stars, which are explained by the oblique rotator model \citep{1950MNRAS.110..395S}. Thus, the observed periodicity of variation is the rotational period of the star. The photometric variability in CP2 stars is thought to result from redistribution of flux in the surface abundance spots \citep[e.g][]{2013A&A...556A..18K}. CP2 stars exhibiting photometric variability are traditionally referred to as $\alpha^2$ Canum Venaticorum (ACV) variables \citep[][GCVS hereafter]{Samu07}.

In the region of the A and F-type stars, where most of the target stars of the present investigation are situated, the $\gamma$ Doradus (GCVS-type GDOR) and $\delta$ Scuti (GCVS-type DSCT) pulsators are found. The GDOR stars exhibit high-order, low-degree, non-radial g-mode pulsations \citep{1999PASP..111..840K} which are driven by convective blocking at the base of the envelope convection zone \citep{2000ApJ...542L..57G}. They were identified as a new class of variable stars by \citet{1994MNRAS.270..905B} and defined as such by \citet{1999PASP..111..840K}. According to the GCVS, they are found between spectral types A7 to F7. Other literature sources narrow this down to A7 to F5 \citep[][and references therein]{2015pust.book.....C}. The DSCT stars, on the other hand, have long been established as a class of variable stars and show low-order g- and p-modes which are self-excited through the $\kappa$ mechanism \citep[e.g.][]{2000ASPC..210....3B}. According to \citet{2015pust.book.....C} and the GCVS, they are mostly found among spectral types A0 to F5. Other sources \citep[e.g.][]{Breg98} have revised the blue border of the DSCT instability strip towards cooler temperatures, corresponding to a spectral type of $\sim$A2, which is what we adopt in the present investigation (cf. Figure 2).

GDOR and DSCT stars are easily distinguished by the timescales of the observed variability. However, the instability strips for both classes partially overlap and hybrid-types are encountered \citep[e.g.][]{2005AJ....129.2026H,2011MNRAS.417..591B}. Different classification criteria are found in the literature. According to the GCVS, GDOR stars are found in the period range of 0.3\,$\leq$\,$P$\,$\leq$\,3 days (0.33\,$\leq$\,$f$\,$\leq$\,3.33 c/d), while DSCT pulsators are characterized by variability in the range of 0.01\,$\leq$\,$P$\,$\leq$\,0.2 days (5\,$\leq$\,$f$\,$\leq$\,100 c/d). \citet{2010ApJ...713L.192G} proposed a division into 'pure' DSCT stars ($f$\,>\,5\,c/d), 'pure' GDOR stars ($f$\,<\,5\,c/d), and hybrid-types exhibiting variability in both frequency regimes. Hybrid-type pulsators are mostly discovered in ultra-precise space photometry but their frequency of occurrence ($\sim$25\% of the sample of \citealt{2011A&A...534A.125U}) has made clear that the picture is complex and the traditional classification scheme might be in need of revision. The understanding and relationship of DSCT and GDOR pulsators is currently in flux (e.g. \citealt{2011A&A...534A.125U}, cf. also \citealt{2015pust.book.....C}).

For many years, the generally held consensus was that -- except in marginal or evolved CP1 stars -- pulsations are absent from classical CP1 stars \citep[e.g.][]{1970ApJ...162..597B,1983spp..book.....W}. In particular, the gravitational settling of helium was proposed to interfere with the $\kappa$ mechanism operating in the He II ionisation zone, thus effectively preventing e.g. DSCT-type pulsations. However, this understanding had to be revoked when \citet{1989MNRAS.238.1077K} identified the classical Am star HD 1097 as a DSCT star.

Several investigations have been carried out to investigate pulsational variability in CP1 stars in the recent past \citep[e.g.][]{2011A&A...535A...3S,2011MNRAS.414..792B,2013MNRAS.429..119P}, which have shown that pulsational variability of GDOR- and DSCT-type is in fact quite common among this group of stars. \citet{2011MNRAS.414..792B} found no significant difference in the location of CP1 DSCT pulsators and non-CP DSCT variables in the Hertzsprung-Russell diagram (HRD). According to \citet{2016arXiv161102254S}, DSCT pulsations in CP1 stars are mostly encountered in the temperature range 6900 $<$ $T_{\rm eff}$ $<$ 7600 K. There is evidence that CP1 pulsators exhibit DSCT pulsations with reduced amplitudes in comparison to normal DSCT stars \citep{2011A&A...535A...3S} but this has not been conclusively established \citep{2016arXiv161102254S}. The results concerning the percentage of pulsators among CP1 stars are partly conflicting (see e.g. the discussion in \citealt{2014PhDT.......131M}). A potential link between chemical peculiarities and hybrid-type pulsation has been proposed \citep{2011ApJ...743..153H}. 

Concerning CP2 stars, the only proven form of pulsational variability is observed in the so-called rapidly oscillating Ap (roAp) stars \citep{1982MNRAS.200..807K}. The possible presence of GDOR and DSCT pulsations in Ap stars have been reported in the literature \citep[e.g.][]{2011MNRAS.410..517B} but this is still a controversial matter. roAp stars exhibit photometric variability with very small amplitudes in the period range of about 5-20 min (high-overtone, low-degree, and non-radial pulsation modes), which is very different from what has been observed for the targets of the present investigation.

It is important to increase the sample size of pulsating CP stars in order to disentangle what mechanisms are at work in these objects, which offer the singular opportunity to study the interaction of atomic diffusion, pulsation, and rotation. The present paper presents an investigation of three CP stars and one CP star candidate which have been identified as hitherto uncatalogued photometrically variable stars during the ongoing search for new ACV variables in public sky survey data \citep{2015A&A...581A.138B,2015AN....336..981B,2016AJ....152..104H}. Newly acquired spectra are presented that confirm the status of our sample stars as chemically peculiar objects, and new photometric observations were acquired in order to verify the results derived from sky survey data. Detailed period analyses have been carried out for all objects. Furthermore, we have derived astrophysical parameters and investigated our sample stars in the $M_{\rm Bol}$ versus $\log T_\mathrm{eff}$ diagram.

Target stars and observations are described in Section \ref{target_stars_and_observations}; data are analysed and results are presented in Section \ref{data_analysis_and_discussion_of_results}. We conclude in Section \ref{conclusions}.

\section{Target Stars and Observations} \label{target_stars_and_observations}

\subsection{Target stars} \label{target_stars}
Except for HD 77013, all objects analysed in this paper are known or suspected CP stars from the Catalogue of Ap, HgMn, and Am stars \citep[][RM09 hereafter]{2009A&A...498..961R} that have been identified as photometrically variable during our efforts in finding new ACV variables \citep{2015A&A...581A.138B,2015AN....336..981B,2016AJ....152..104H}. HD 77013 is a late A-type star that was included as a candidate ACV variable in \citet{2015A&A...581A.138B}.

Table 1 provides essential data for the target stars of the present investigation, which are listed in order of increasing right ascension. Positional information was taken from the Tycho-2 catalogue \citep{2000A&A...355L..27H}. The $(B-V)$ and $(J-K\textsubscript{s})$ indices\footnote{According to the 2MASS catalogue, the $J$ magnitude of HD 98000 has been based on poor photometry. However, the observed value agrees within the errors with the corresponding $J$ magnitude from the DENIS survey \citep{1997Msngr..87...27E}.} were taken from \citet{2001KFNT...17..409K} and \citet{2006AJ....131.1163S}, respectively. The $V$ magnitude range was derived from ROAD (HD 67983; see section \ref{new_ccd_photometry} for a description of the ROAD data) and ASAS-3 (all other objects) $V$-band data. Spectral types given were taken from RM09 (HD 67983, HD 81076, HD 98000), and \citet[][HD 77013]{1999mctd.book.....H}. As in the original catalogue, the 'p' denoting peculiarity has been omitted from the spectral classifications taken from RM09.

\begin{table}
	\centering
	\caption{Essential data for the target stars of the present investigation.}
	\label{table1}
	\begin{adjustbox}{max width=0.5\textwidth}
	\begin{tabular}{lllll}
		\hline
								&	HD 67983 & HD 77013	& HD 81076 & HD 98000 \\
		\hline
		ID RM09	& 18780	& n/a	& 23020	& 28280 \\
		RA (J2000) & 08 08 22.960	& 08 59 34.002 & 09 22 03.971	& 11 15 36.775 \\
		Dec (J2000)	& -46 36 33.18 & -05 52 09.54	& -45 38 34.37 & -53 03 43.60 \\
		Range (V)	& 9.55 -- 9.63 & 9.23 -- 9.29	& 9.58 -- 9.62	& 10.47 -- 10.51 \\
		Sp.type (lit) & F0- dD? & A9	& A2 Eu Cr (?) & B9 Si \\
		$(B-V)$ ind. & 0.383 & 0.323 & 0.231 & 0.068 \\
		$(J-K\textsubscript{s})$ ind.	& 0.058	& 0.157	& 0.145	& -0.022 \\
		\hline
	\end{tabular}
	\end{adjustbox}
\end{table}

\subsection{Archival Time Series Photometry} \label{archival_time_series_photometry}
All of our targets have been discovered as photometrically variable stars using data from the third phase of the ASAS \citep[ASAS-3;][]{2002AcA....52..397P} and SuperWASP \citep[SWASP;][]{2010A&A...520L..10B} projects. In total, 2813 ASAS-3 measurements (all targets) and 9401 SWASP measurements (HD 81076) were included into the analysis. Below is given a short description of both data sources, which are perfectly suited for an investigation of the low-amplitude photometric variability of chemically peculiar stars \citep[cf. also][]{2015A&A...581A.138B,2015AN....336..981B}. No data are available for our target stars in the CoRoT \citep{2009A&A...506..411A} and Kepler \citep{2010Sci...327..977B} databases.

\subsubsection{Characteristics of ASAS-3 data} \label{characteristics_of_ASAS-3_data}
The All Sky Automated Survey (ASAS) aims at constant photometric monitoring of the whole available sky, with the ultimate goal of detecting and investigating any kind of photometric variability. The third phase of the project, ASAS-3, lasted from 2000 until 2009 \citep{2002AcA....52..397P} and has been monitoring the entire southern sky and part of the northern sky ($\delta$\,$<$\,+28\degr). The ASAS-3 system, situated at the 10-inch astrograph dome of the Las Campanas Observatory in Chile, comprised two wide-field telescopes equipped with f/2.8 200\,mm Minolta lenses and 2048 x 2048 AP 10 Apogee detectors, covering a field of 8\fdg8x8\fdg8 of sky. Data were obtained in the Johnson $V$ passband for about 10$^{7}$ sources brighter than $V \approx$ 14 mag. CCD resolution is about 14\farcs8 / pixel, which results in an astrometric accuracy of around 3 -- 5\arcsec\ for bright stars and up to 15.5\arcsec\ for fainter stars \citep{2014IAUS..301...31P}. This leads to uncertain photometry in crowded fields.

The ASAS-3 archive contains reasonable photometry for stars in the magnitude range 7\,$\leq$\,$V$\,$\leq$\,14; the most accurate photometry was obtained for stars in the magnitude range 8\,$\la$\,$V$\,$\la$\,10, boasting a typical scatter of about 0.01 mag \citep[e.g.][]{2014IAUS..301...31P}. Because of the long time-baseline of almost ten years, though, the detection limit in ASAS-3 data for periodic signals is much lower. \citet{2014JAD....20....5D} detected periodic variables with a range of variability of 0.01--0.02 mag in the magnitude range of 7\,$\la$\,$V$\,$\la$\,10. \citet{2014IAUS..301...31P} estimated that periodic signals with amplitudes as low as about 5\,millimag (mmag) can be detected. 

The all-sky character of the survey necessitated sparse sampling; generally, a field was observed every one, two or three days \citep{2014IAUS..301...31P}, which results in strong daily aliases in Fourier amplitude spectra.

\subsubsection{Characteristics of SuperWASP data} \label{characteristics_of_SuperWASP_data}
The SuperWASP project mainly aims at detecting transiting extrasolar planets and employs two robotic telescopes, which are situated at the Observatorio del Roque de los Muchachos (La Palma) and the South African Astronomical Observatory (SAAO). At each site, an array of eight f/1.8 200\,mm Canon lenses and 2048 x 2048 Andor CCD detectors is arranged on a single equatorial fork mount, covering a field of 7\fdg8x7\fdg8 at a plate scale of 13.7\arcsec\ / pixel \citep{2006PASP..118.1407P}. Observations consist in general of two consecutive 30s integrations; a field was typically observed about every 9 to 12 min \citep{2010A&A...520L..10B}.
The first data release (DR1) of the SuperWASP project covers a large fraction of the sky and encompasses photometric observations from 2004 to 2008. It boasts $\sim$18 million light curves and provides photometry in a broadband filter (4000-7000 \AA) with an accuracy better than 1\% for objects in the magnitude range 8\,$\leq$\,$V$\,$\leq$\,11.5 mag \citep{2006PASP..118.1407P}.

\subsection{New CCD Photometry} \label{new_ccd_photometry}
In order to verify the results from the sky surveys, new CCD photometric observations of our targets were acquired at the Remote Observatory Atacama Desert \citep[ROAD;][]{Hamb12}. Observations have been taken with an Orion Optics, UK Optimized Dall Kirkham 406/6.8 telescope and a FLI 16803 CCD camera. Observations were taken through a clear filter (HD 98000, in order to maximise throughput) and an Astrodon Photometric $V$ filter (all other objects).

With 8s exposure time in the 3x3 binning mode, a total of 4359 measurements were acquired. Twilight sky-flat images were used for flatfield corrections. The reductions were performed with the MAXIM DL program\footnote{http://www.cyanogen.com} and the determination of magnitudes using the LesvePhotometry program\footnote{http://www.dppobservatory.net/}.

\subsection{Spectroscopic Observations} \label{spectroscopic_observations}
\begin{figure}
\includegraphics[width=\columnwidth]{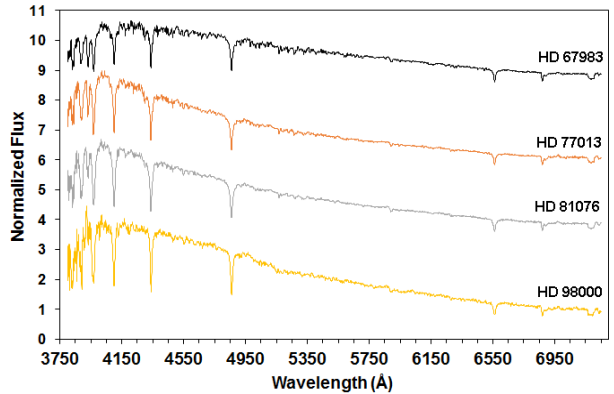}
\caption{An overview of the classification resolution spectra of our target stars, obtained at Mirranook Observatory. Details of instrumentation are provided in the text. The spectra have been offset by an arbitrary amount for clarity.}
\label{spectra_overview}
\end{figure}

Spectra of our target stars were obtained at Mirranook Observatory using a LISA spectrograph with a 24 $\micron$ slit on a C11 279/2800 Schmidt-Cassegrain telescope and an Atik 314+ camera (S/N$\sim$60; R$\sim$1500). With the same setup, spectra of MK standards from the lists of \citet{Gray89a, Gray89b} and well-established CP1 stars from the RM09 catalogue were taken. Additional spectra were acquired with a Spectra L200 Littrow spectroscope on a 0.25m Schmidt-Cassegrain telescope and an Atik 383L+ camera (S/N$\sim$70; R$\sim$2000). The spectra are instrument corrected using a Miles standard star taken at similar airmass and were processed with the ISIS software package \citep{2013LPI....44.2318A}. An overview of the newly acquired spectra is given in Figure \ref{spectra_overview}.

The resolution $R$ of our spectra corresponds to a line broadening of about 100 to 125 km\,s$^{-1}$ according to the formula ${\upsilon}\sin i \leq \frac{c}{1.5R}$ 
\citep{Jasch90}. It is not possible to resolve projected rotational velocities lower than these values.

All programme stars were classified in the framework of a refined Morgan-Keenan-Kellman system (MKK hereafter), as described in \citet{Gray87, Gray89a, Gray89b}, using the standard techniques of MKK classification. For a precise classification of the objects and to identify possible peculiarities, the spectra were compared visually and overlaid with MKK standards. In addition, we used the spectra of well-established CP stars to guide us in the identification of the various peculiarities. Whenever possible, the comparison spectra were obtained at Mirranook Observatory with the same instrumentation as the spectra of our programme stars (HD 26690, HD 29499, HD 40136); the other spectra were taken from \citet{2001A&A...373..625P} (HD 23194, HD 23643, HD 58946). The detection of spectral peculiarities is straightforward with the achieved signal-to-noise ratio, whereas the error for the spectral classification is about $\pm$1\,subclass \citep{Pa01}.

The resolution of our spectra is rather low, and lines which are important in CP1 stars (like e.g. Sc II $\lambda$4246 \AA) appear blended with other nearby spectral features. However, investigating and comparing such conspicuous features as the Ca II K line, the hydrogen-line strengths and the general metallic-line spectrum is a straightforward task and should not be impeded by the low resolution of our spectroscopic material. CP1 stars have been traditionally classified on spectra with plate factors between 40 and 120 \AA/mm, and may even be singled out at 300 \AA/mm \citep{Jasch90}. We therefore think that our spectra are sufficient for a first investigation of the chemical composition of our sample stars. High-resolution spectroscopy would be of benefit in reducing the remaining uncertainties and in deriving precise atmospheric parameters, e.g. calcium, scandium and iron-peak element abundances, which are important indicators of CP1 stars. On the background of the proposed pulsational variability, these findings would be of great astrophysical interest \citep[e.g.][]{2015MNRAS.451..184C}. On these grounds, we intend to secure high-resolution spectroscopic observations of our target stars, which will be presented in a forthcoming investigation.

The results of our spectral classification are provided in Table \ref{data}, together with other astrophysical parameters derived in Section \ref{HR_diagram}. A comparison with MK standard spectra from the lists of \citet{Gray89a, Gray89b} and a discussion of the individual spectra and the derived spectral types is presented in Section \ref{discussion_of_individual_objects}.

\section{Data Analysis and Discussion of Results} \label{data_analysis_and_discussion_of_results}

\subsection{Data Analysis} \label{data_analysis}

Data of our target stars were downloaded from the ASAS-3 website\footnote{http://www.astrouw.edu.pl/asas/} and the SuperWASP archive at the computing and storage facilities of the CERIT Scientific Cloud\footnote{http://wasp.cerit-sc.cz/} \citep{2014IBVS.6090....1P}. All light curves were inspected visually; obvious outliers and data points associated to exceedingly large error bars were removed. Care was taken in this step as the removal of only a few datapoints may have a profound impact on the derived frequencies. Furthermore, data points with a quality flag of 'D' (= 'worst data, probably useless') were deleted from the ASAS-3 datasets. No long-term trends were identified in the ASAS datasets; further details concerning the processing of the SWASP data can be found in section \ref{HD_81076_discussion}.

The period analysis was done using PERIOD04 \citep{2005CoAst.146...53L}. In order to extract all relevant frequencies, the data were searched for periodic signals and consecutively prewhitened with the most significant frequency. The detailed frequency analyses are presented in the discussion of the corresponding objects (cf. section \ref{discussion_of_individual_objects}).

Generally, ROAD measurements are of higher accuracy than ASAS-3 observations. However, this intrinsic disadvantage is amply compensated for by the ASAS survey's time base of almost ten years, which renders the detection of periodic signals with very small amplitudes possible (down to about 5 mmag; cf. section \ref{characteristics_of_ASAS-3_data}). We estimate the threshold of detection of periodic signals to be about 3--4 mmag for ROAD data; thus, both datasets are comparable in this respect.

Because the time base of the ASAS-3 survey is longer by several orders of magnitude in comparison to the ROAD measurements, ASAS data boast better frequency resolution and are more suited to the disentangling of multiple frequencies and will yield more accurate results. Thus, they have always been given more weight in the interpretation of results in the following sections.

SWASP data are available for only one of our target stars. Concerning accuracy and observing cadence, these data are superior to both other datasets. Several studies have shown that signals with an amplitude of $\sim$1 mmag are still detectable in data from the SWASP project \citep[e.g.][]{2011A&A...535A...3S}.

Aliasing is a problem affecting the results from both datasets, in particular the ASAS data. Generally, $(f\textsubscript{i} \pm1)$ c/d should be considered as possible frequencies. However, as the derived best fitting frequencies are in almost all cases confirmed in every available dataset, we are very confident of their significance. Details concerning the frequency solution for our target stars are presented in the discussion of the individual objects (section \ref{discussion_of_individual_objects}). Generally, different mechanisms like e.g. rotation or pulsation are at work in A- and F-type stars; therefore, unravelling the cause of the observed variability is not a trivial task.

\subsection{Astrophysical Parameters and the $M_{\rm Bol}$ versus $\log T_\mathrm{eff}$ diagram} \label{HR_diagram}

Photometric data were taken from the General Catalogue of Photometric Data \citep[GCPD,][]{Merm97}\footnote{http://gcpd.physics.muni.cz/}. Where possible, averaged and
weighted mean values were used throughout.

The reddening was estimated using photometric calibrations in the Str{\"o}mgren $uvby\beta$ system \citep{Craw78,Craw79,Domi99} and the interstellar extinction models by 
\citet{Arenou92}, \citet{Chen98}, and \citet{Fitzgerald68}. The agreement of all these estimates is excellent.

For the estimation of $T_\mathrm{eff}$, the following calibrations were used:

\begin{itemize}
\item Geneva system: \citet{Cram84,Cram99}, \citet{Kuenz97}, based on the reddening free parameters $X$ and $Y$ which are valid for spectral types hotter than approximately A0. The results are therefore independent of the estimation of $A_V$ for HD 98000. For stars cooler than 8500\,K, \citet{Kuenz97} used the grids of $m_2$ and $d$
versus $(B2-V1)_0$ which served as a calibration tablet. The definition of these indices are listed in \citet{Golay94}.
\item Str{\"o}mgren system: \citet{na93}, based on the unreddened $(u-b)$ colour for HD 98000 and the $T_\mathrm{eff}$ versus $(b-y)_0$ relation (Eq. 10 therein), which was applied to the other three stars.
\end{itemize}

In addition, we applied the effective temperature calibration by \citet{net07}, which was especially developed for the different CP star subgroups (see Table \ref{data} for
the subgroups of our targets). The individual effective temperature values within each photometric system were first checked for their intrinsic consistency and then averaged.

The $M_{\rm Bol}$ values were directly calculated from the astrometric parallaxes of the Gaia DR1 \citep{2016arXiv160904172G,2016arXiv160904303L}. The bolometric corrections were taken from \citet{net07} and an absolute bolometric magnitude of the Sun of $M_{\rm Bol}(\odot)$\,=\,4.74\,mag was used. The error of the bolometric magnitude was directly deduced from the error of the parallax.

Characteristics and astrophysical parameters of our sample stars are listed in Table \ref{data}. The errors in the final digits of the corresponding quantity are given in parentheses.

\begin{figure}
\begin{center}
\includegraphics[width=\columnwidth]{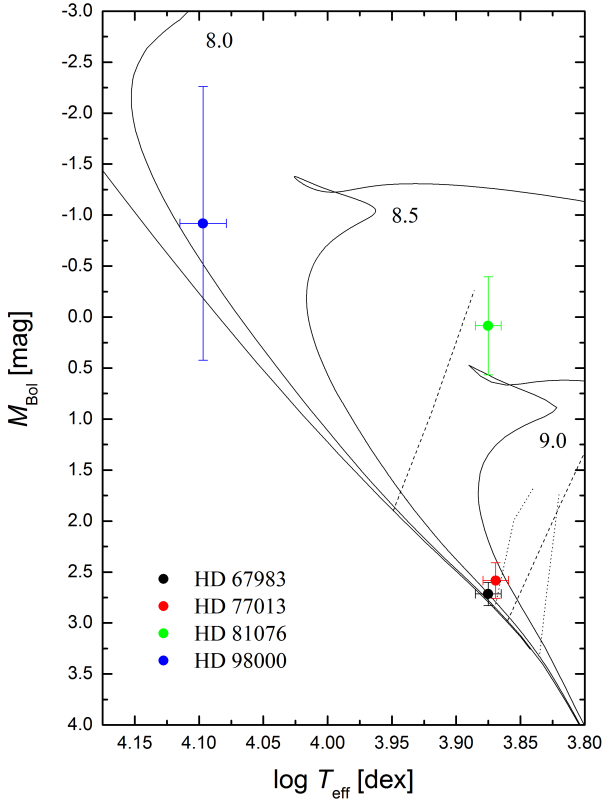}
\caption{The $M_{\rm Bol}$ versus $\log T_\mathrm{eff}$ diagram of our programme stars (Table \ref{data}). The borders of the $\delta$\,Scuti (dashed lines) and $\gamma$\,Doradus (dotted lines) instability strips are indicated, which were taken from \citet{Breg98} and \citet{Dupr04}, respectively. The isochrones are from the PARSEC database \citep{Bress12} for solar metallicity ([Z]\,=\,0.019\,dex).}
\label{hrd} 
\end{center} 
\end{figure}

Figure \ref{hrd} shows the location of the target stars in the $M_{\rm Bol}$ versus $\log T_\mathrm{eff}$ diagram. Also indicated are PARSEC isochrones \citep{Bress12} for solar metallicity ([Z]\,=\,0.019\,dex) and the borders of the $\delta$\,Scuti as well as $\gamma$\,Doradus instability strips taken from \citet{Breg98} and \citet{Dupr04}, respectively. 

From this figure, we conclude:

{\it HD 67983 and HD 77013:} Both stars are very similar, which is reflected in their positions in the $M_{\rm Bol}$ versus $\log T_\mathrm{eff}$ diagram. They are both on, or very close to, the Zero Age Main Sequence and are situated near the blue border of the $\gamma$\,Doradus instability strip, in the region where the $\delta$\,Scuti and $\gamma$\,Doradus instability strips overlap. These objects are good candidates for showing both GDOR and DSCT-type pulsations.

{\it HD 81076:} This star seems to be evolved beyond the Terminal Age Main Sequence and is located within the classical $\delta$\,Scuti instability strip. We estimate an age between 600 Myr and 1 Gyr.

{\it HD 98000:} Classical pulsation can be excluded as the cause of light variation because of the star's high effective temperature. Unfortunately, the large error of the parallax prevents estimating the evolutionary status and thus its age and mass with high precision. Further spectrophotometric data are needed to shed more light on the star's astrophysical parameters. 

\begin{table}
\begin{center}
\caption{Characteristics and astrophysical parameters of our sample stars.}
\label{data}
\begin{adjustbox}{max width=0.5\textwidth}
\begin{tabular}{ccccccc}
\hline
\hline
ID & $\pi$ & $A_V$ & $\log T_\mathrm{eff}$ & $M_{\rm Bol}$ & Spectral & CP \\
& [mas] & [mag] & [dex] & [mag] &	type &	subgroup \\       
\hline
HD 67983	&	4.76(25)	&	0.09	&	3.875(9)	&	+2.71(11)	&	kF0hF0mF3\,V &	CP1(Am:)	\\
HD 77013	&	4.95(40)	&	0.12	&	3.869(9)	&	+2.58(18)	&	kA5hA9mF2\,V &	CP1	\\
HD 81076	&	1.31(29)	&	0.07	&	3.875(9)	&	+0.09(48)	&	kA3hA8mF1\,V &	CP1 \\
HD 98000	&	0.76(47)	&	0.24	&	4.097(17)	&	$-$0.91(1.34) 	& B9pSi &	CP2 \\
\hline
\end{tabular}
\end{adjustbox}
\end{center}
\end{table}

\subsection{Discussion of Individual Objects} \label{discussion_of_individual_objects}

\subsubsection{HD 67983} \label{HD_67983_discussion}
HD 67983 was initially classified as spectral type F0 in the Henry Draper Catalogue \citep{1919AnHar..93....1C} and later identified as a metallic-line star ('Fm Delta Del or late Am') by \citet{1978mcts.book.....H}. On this basis, it was included in the RM09 catalogue which lists a spectral type of 'F0- dD?'. The star has never been subjected to a variability study before.

\begin{figure}
	\includegraphics[width=\columnwidth]{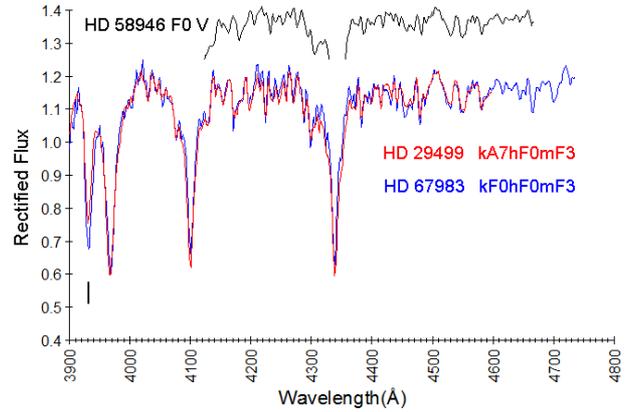}
    \caption{A comparison of the blue-violet spectrum (3800--4600 \AA) of HD 67983 (kF0hF0mF3V; blue) to the spectrum of the well-established CP1 star HD 29499 (kA7hF0mF3; red). For comparsion, part of the spectrum of the F0V MK standard HD 58946 (black) is also shown. Note the difference in Ca II K line strength between HD 67983 and HD 29499. Details on the spectroscopic observations are given in section \ref{spectroscopic_observations}.}
    \label{fig_HD67983_Am}
\end{figure}

A comparison of the blue-violet spectrum (3800--4600 \AA) of HD 67983 (kF0hF0mF3V; blue) to the spectrum of the well-established CP1 star HD 29499 is shown in Figure \ref{fig_HD67983_Am}. HD 29499 is listed with Sp(K)\,=\,A7 and Sp(m)\,=\,F3 in the RM09 catalogue. There is a good match between the hydrogen-line and metallic-line strengths of both stars. For comparison, part of the spectrum of the F0V MK standard HD 58946 is also shown, from which becomes clear that the metallic-line spectrum is decidedly later than F0 in both stars. However, the Ca II K line of HD 67983 is not as weak as in the classical Am star HD 29499. We therefore conclude that HD 67983 is a marginal CP1 (Am:) star and derive a spectral type of kF0hF0mF3V.

\begin{figure}
	\includegraphics[width=\columnwidth]{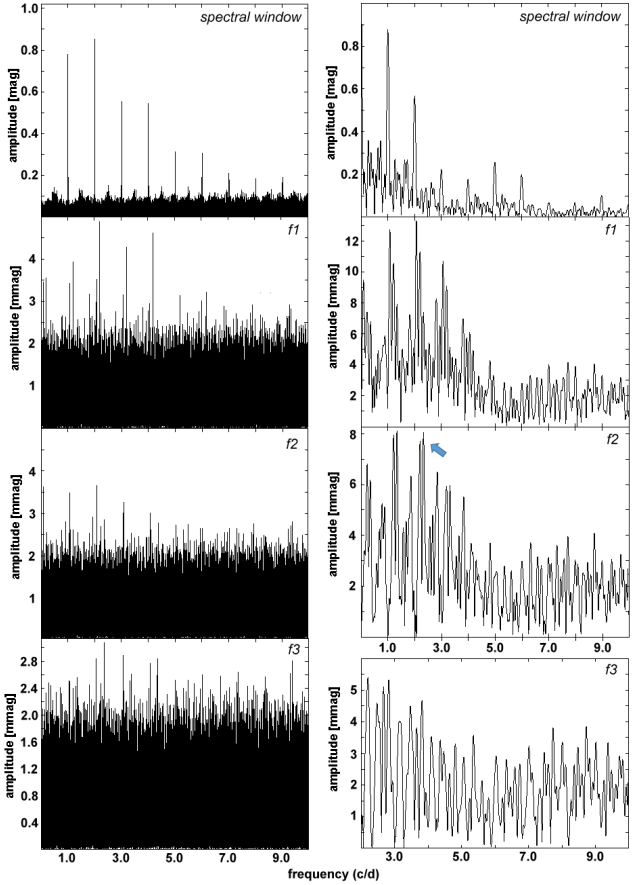}
    \caption{Period analysis of ASAS (left panels) and ROAD data (right panels) for HD 67983, illustrating the different steps of the frequency spectrum analysis. The top panels show the spectral windows dominated by daily aliases. Middle and bottom panels show the frequency spectra for unwhitened data and data that has been prewhitened with $f1$ and $(f1+f2)$, respectively. Note the different scales on the ordinates and the abscissa in the bottom panel on the right-hand side of the plot.}
    \label{fig_HD67983_pa}
\end{figure}

We have analysed photometric data from the ASAS-3 archive (608 observations; time base of $\sim$3300 days) and our own observations (794 observations; time base of $\sim$11.3 days). The results are presented in Figure \ref{fig_HD67983_pa} and Table \ref{HD67983_table}. The numbers in parentheses indicate the errors in the final digits. Error margins were estimated using Period04 and employing the width of the corresponding peaks in the power spectra.

Due to the low angular resolution of the ASAS instruments (cf. section \ref{characteristics_of_ASAS-3_data}), ASAS-3 measurements for HD 67983 are slightly affected by light contribution from the nearby ($\sim$15.5\arcsec\ distant) UCAC4 217-019803 ($V$\,=\,15.13 mag), which might explain the relatively large difference in zero point between the ASAS-3 and our own $V$ band measurements and result in a reduced amplitude of variability in the ASAS-3 dataset.

\begin{table}
\begin{center}
\caption{HD 67983; frequencies and corresponding semi-amplitudes detected in the ROAD and ASAS-3 datasets, as derived with PERIOD04.}
\label{HD67983_table}
\begin{adjustbox}{max width=0.5\textwidth}
\begin{tabular}{ccc}
\hline
\hline
ID & frequency & semi-amplitude \\
   & [c/d]     & [mag] \\       
\hline
$f1\textsubscript{ASAS}$	& 2.20201(2) & 0.0051 \\
$f2\textsubscript{ASAS}$	& 2.08179(2) & 0.0039 \\
$f3\textsubscript{ASAS}$  & 2.35871(3) & 0.0032 \\
\hline
$f1\textsubscript{ROAD}$	& 2.064(4) & 0.0116 \\
$f2\textsubscript{ROAD}$	& 1.333(7) / 2.333(7)$^{1}$ & 0.0068 \\
$f3\textsubscript{ROAD}$	& 2.204(7) & 0.0064 \\
\hline
\multicolumn{3}{l}{$^{1}$ See text for details.}
\end{tabular}
\end{adjustbox}
\end{center}
\end{table}

Fig. \ref{fig_HD67983_pa} illustrates that the derived frequency spectra are strongly affected by daily aliases, which is to be expected when analysing data from single-site observations. This adds ambiguity to the frequency identification and, as has been pointed out in section \ref{data_analysis}, it cannot be excluded that the derived frequencies are aliases of the true values.

Three closely-spaced frequencies have been derived from ASAS-3 data, although the third peak is of rather low amplitude and signal to noise. Within the errors, all frequencies are recovered in ROAD data, although -- interestingly -- with significantly different amplitudes. The dominant frequency in ROAD data corresponds to the frequency with the second highest amplitude in the ASAS-3 dataset ($f1\textsubscript{ROAD}$ $\sim$ $f2\textsubscript{ASAS}$). After prewhitening for $f1$, a frequency of 1.33\,c/d is derived from ROAD data, which is likely the $(f-1)$ c/d alias of the third frequency detected in the ASAS-3 data ($f3\textsubscript{ASAS}$). As ASAS-3 data are much more suited for the disentangling of multiple frequencies, and the peak near 2.33\,c/d is of comparable power in ROAD data, too (cf. Fig. \ref{fig_HD67983_pa}), we are inclined to interpret the 1.33\,c/d frequency in ROAD data as an alias of the true period and have therefore adopted the frequency near 2.33\,c/d (marked by the blue arrow in Fig. \ref{fig_HD67983_pa}). Restricting the period search to the region 2\,<\,$f(c/d)$\,<\,10 in order to exclude the low-frequency noise present in this ROAD dataset, we next come up with a frequency of 2.20\,c/d, which corresponds to $f1\textsubscript{ASAS}$. No further significant frequencies could be extracted.

\begin{figure}
	\includegraphics[width=\columnwidth]{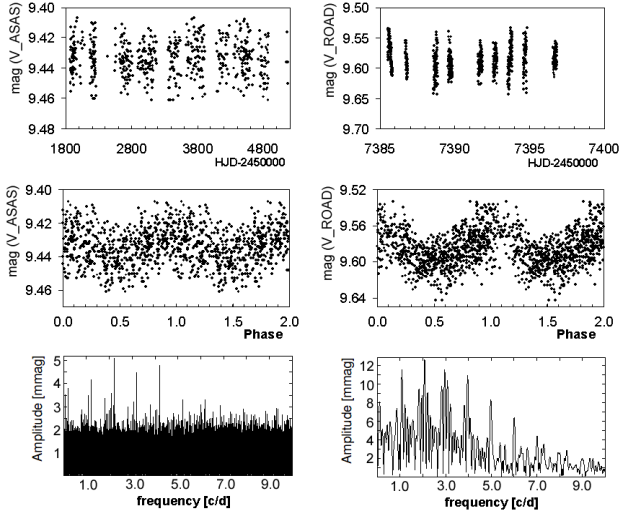}
    \caption{Light curves, phase plots and unwhitened frequency spectra of HD 67983, based on ASAS-3 data (left panels) and ROAD data (right panels).}
    \label{fig_HD67983_pp}
\end{figure}

As can be judged from Fig. \ref{fig_HD67983_pa}, the third frequencies in both ASAS-3 and ROAD data are of low amplitude and signal to noise. They are interpreted here as significant features on grounds of their confirmation in both datasets but need independent confirmation. Data distribution and phase plots, folded with the derived best fitting frequencies, are illustrated in Fig. \ref{fig_HD67983_pp}.

The difference in time base and resolution might go a long way in explaining the observed differences between both datasets. However, the agreement between the results derived from ASAS-3 and ROAD data are excellent for all other stars of this investigation. Therefore, the possibility should be considered that the observed discrepancies might have a physical background, i.e. are intrinsic to the variability pattern of the star. For example, long-term amplitude changes of pulsational modes have been observed \citep[e.g.][]{2008A&A...477..907P}. Following this scenario, the frequency near $\sim$2.06\,c/d might have been the most prominent one during the $\sim$11 days of our photometric coverage, whereas its amplitude had been averaged out over the longer time base of the ASAS-3 dataset. Another explanation might be the influence of the beating of unresolved frequencies, which are beyond the detection limit of both datasets. Only further (multisite) observations will be able to settle this matter conclusively.

The observed, closely-spaced frequencies, the complex variability pattern (e.g. the proposed changes in amplitude) and the absence of harmonics, which are indicative of localised spots and a characteristic of the frequency spectra of rotating variables \citep[e.g.][]{2015MNRAS.451.1445B}, are in agreement with pulsational variability and provide evidence against rotational modulation as the cause of the observed brightness changes. Furthermore, rotationally induced variability in Am stars is still a matter of debate. The results of \citet{2015MNRAS.448.1378B} suggest the existence of rotational modulation in this group of stars; however, with observed amplitudes of up to $\sim$200 ppm, this kind of variability is reserved for high-precision (space) photometry and very different from what has been observed for HD 67983.

Therefore, on the basis of the available data, we consider pulsational variability as the most likely cause of the photometric variability of the marginal CP1 (Am:) star HD 67983. The derived frequencies and amplitudes fall inside the realm of the GDOR variables, and the observed spectral type (kF0hF0mF3V) is also in agreement with a GDOR classification. This is further corroborated by the star's position in the $M_{\rm Bol}$ versus $\log T_\mathrm{eff}$ diagram (cf. Figure \ref{hrd}). We therefore propose HD 67983 as a new GDOR variable, thus enlarging the sample of CP objects among this class of variable stars \citep{2011A&A...535A...3S}.

\subsubsection{HD 77013} \label{HD_77013_discussion}
HD 77013 was classified as spectral type A3 in the Henry Draper Catalogue \citep{1919AnHar..93....1C} and later reclassified as A9 by \citet{1999mctd.book.....H}. No chemical peculiarities of HD 77013 have been reported in the literature. However, the star was identified as a promising ACV variable candidate on grounds of its typical photometric variability by \citet{2015A&A...581A.138B}, who derived a period of $P$\,=\,0.411738\,d ($f$\,=\,2.428728\,c/d) from an analysis of ASAS-3 data. The star has never been subjected to a detailed study before.

\begin{figure}
	\includegraphics[width=\columnwidth]{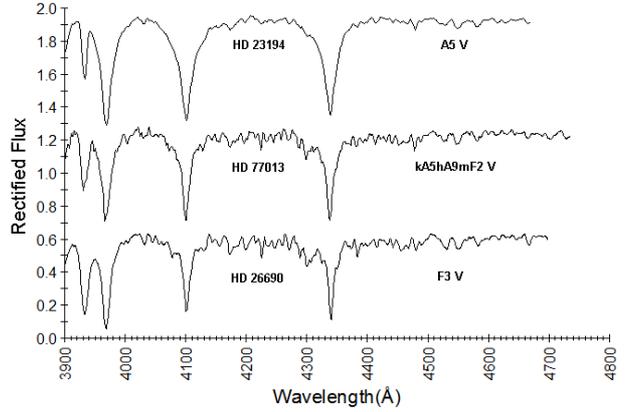}
    \caption{A comparison of the blue-violet spectrum (3800--4600 \AA) of HD 77013 (kA5hA9mF2V) to the spectra of the two MK standard stars HD 23194 (A5V) and HD 26690 (F3V). Details on the spectroscopic observations are given in section \ref{spectroscopic_observations}.}
    \label{fig_HD77013_Am}
\end{figure}

A comparison of the blue-violet spectrum (3800--4600 \AA) of HD 77013 to the spectra of the two MK standard stars HD 23194 (A5V) and HD 26690 (F3V) is illustrated in Figure \ref{fig_HD77013_Am}. It can be clearly seen that the Ca II K line of HD 77013 is similar in strength to that of the A5V MK standard. The metallic-line spectrum, however, is much more pronounced in HD 77013, being slightly earlier than that of the F3V MK standard star HD 26690. The hydrogen-line spectrum indicates an intermediate type which we estimate to about A9. We deduce that HD 77013 is a classical CP1 star and derive a spectral type of kA5hA9mF2V.

\begin{figure}
	\includegraphics[width=\columnwidth]{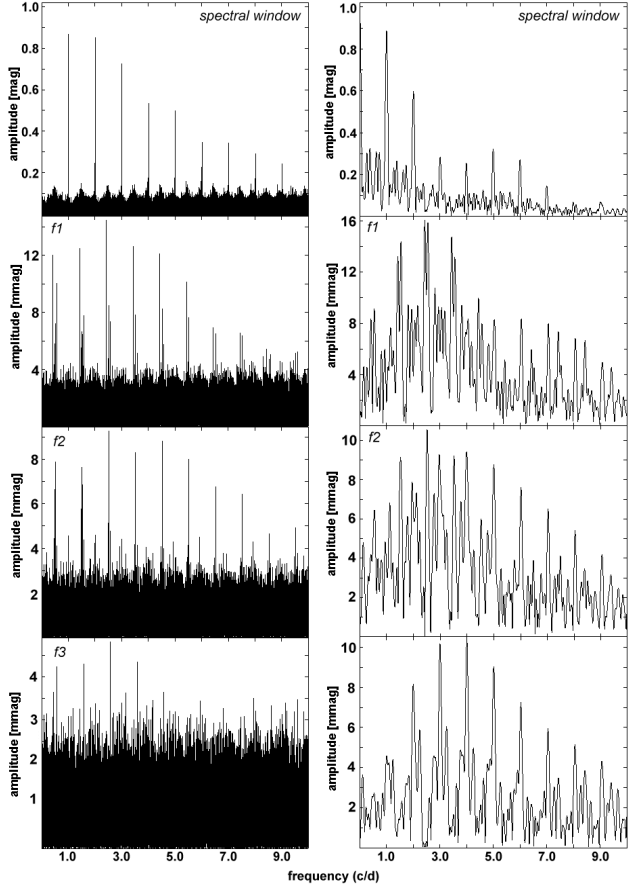}
    \caption{Period analysis of ASAS (left panels) and ROAD data (right panels) for HD 77013, illustrating the different steps of the frequency spectrum analysis. The top panels show the spectral windows dominated by daily aliases. Middle and bottom panels show the frequency spectra for unwhitened data and data that has been prewhitened with $f1$ and $(f1+f2)$, respectively. Note the different scales on the ordinates.}
    \label{fig_HD77013_pa}
\end{figure}

We have analysed photometric data from the ASAS-3 archive (549 observations; time base of $\sim$3300 days) and our own observations (693 observations; time base of $\sim$11.1 days). We have checked the corresponding sky region of HD 77013 using the ALADIN interactive sky atlas \citep{2000A&AS..143...33B}; no star of sufficient brightness is situated near our target to affect the ASAS measurements. The results of our analysis are given in Figure \ref{fig_HD77013_pa} and Table \ref{HD77013_table}.

Analysis of ASAS-3 data reveals three closely-spaced frequencies, which -- judging from their amplitudes and the structure of the Fourier spectra -- all seem relevant. Only the first two frequencies are recovered in ROAD data; a third frequency is very close to 4 c/d (see Fig. \ref{fig_HD77013_pa}, bottom right panel) and thus likely an intrinsic characteristic of the dataset.

\begin{figure}
	\includegraphics[width=\columnwidth]{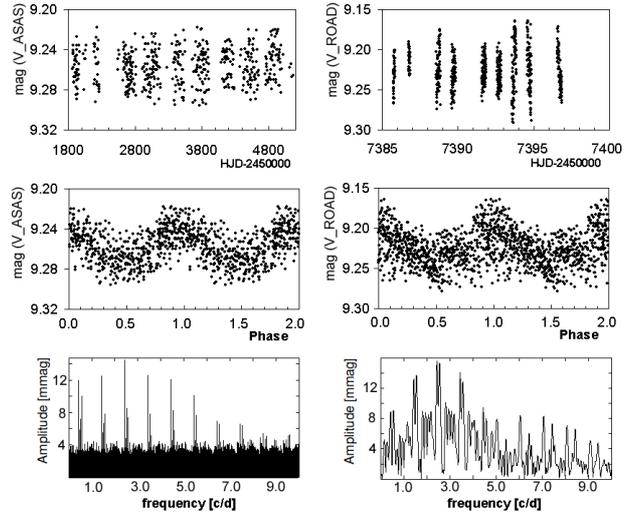}
    \caption{Light curves, phase plots and unwhitened frequency spectra of HD 77013, based on ASAS-3 data (left panels) and ROAD data (right panels).}
    \label{fig_HD77013_pp}
\end{figure}

The strongly daily aliasing inherent to the ASAS-3 and ROAD datasets is apparent from Fig. \ref{fig_HD77013_pa}. However, the two most significant frequencies in both datasets agree within the errors ($f1\textsubscript{ROAD}$ $\sim$ $f1\textsubscript{ASAS}$; $f2\textsubscript{ROAD}$ $\sim$ $f2\textsubscript{ASAS}$), which makes us confident of their reality and significance. We therefore consider the peaks corresponding to $(f\textsubscript{i} \pm1)$ c/d to be 1-day aliases. Data distribution and phase plots, folded with the derived best fitting frequencies, are illustrated in Fig. \ref{fig_HD77013_pp}.

\begin{table}
\begin{center}
\caption{HD 77013; frequencies and corresponding semi-amplitudes detected in the ROAD and ASAS-3 datasets, as derived with PERIOD04.}
\label{HD77013_table}
\begin{adjustbox}{max width=0.5\textwidth}
\begin{tabular}{ccc}
\hline
\hline
ID & frequency & semi-amplitude \\
   & [c/d]     & [mag] \\       
\hline
$f1\textsubscript{ASAS}$	& 2.42872(1) & 0.0148 \\
$f2\textsubscript{ASAS}$	& 2.52219(1) & 0.0090 \\
$f3\textsubscript{ASAS}$  & 2.57618(2) & 0.0049 \\
\hline
$f1\textsubscript{ROAD}$	& 2.435(4) & 0.0132 \\
$f2\textsubscript{ROAD}$	& 2.525(4) & 0.0121 \\
\hline
\end{tabular}
\end{adjustbox}
\end{center}
\end{table}

The closely-spaced frequencies found in the ASAS-3 dataset are reminiscent of the phenomenon of rotational splitting of non-radial oscillation modes \citep[e.g.][]{2010ApJ...721.1900D}. No equidistant triplet of frequencies is observed for HD 77013, though ($f2-f1$ $\sim$ 0.0935; $f3-f2$ $\sim$ 0.0540). Theoretically, differential rotation might also result in multiple, closely-spaced frequency peaks \citep[e.g.][]{2013A&A...560A...4R}, although this phenomenon is usually associated with star spots and not to be expected in an early-type main sequence star. However, Balona and coworkers have recently collected evidence that A-type stars may be active and show starspots in the same way as their cooler counterparts \citep[e.g.][]{2013MNRAS.431.2240B}. Nevertheless, this is still a matter of controversy \citep[e.g.][]{2014PhDT.......131M} and the observed amplitudes are very small, in contrast to what has been observed for HD 77013. This scenario seems therefore unsuited to explain the observed variability.

In the light of the above mentioned probabilities, we propose pulsational variability as the most likely cause of the observed photometric variability of HD 77013. Considering its spectral type (kA5hA9mF2V), position in the $M_{\rm Bol}$ versus $\log T_\mathrm{eff}$ diagram (cf. Figure \ref{hrd}), and the observed time scale of variability, HD 77013 is most likely a CP1 GDOR pulsator.

\subsubsection{HD 81076} \label{HD_81076_discussion}
Originally classified as spectral type A2 in the Henry Draper Catalogue \citep{1919AnHar..94....1C}, HD 81076 was later reclassified as 'ApEuCr, weak case' by \citet{1978mcts.book.....H}. Consequently, the star is listed in the RM09 catalogue with a spectral type of A2pEuCr and a remark denoting the doubtful nature of its status as a CP star. HD 81076 has never been the subject of a variability study before.

\begin{figure}
	\includegraphics[width=\columnwidth]{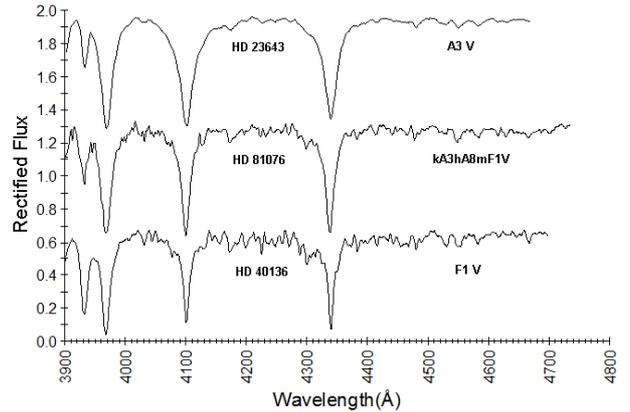}
    \caption{A comparison of the blue-violet spectrum (3800--4600 \AA) of HD 81076 (kA3hA8mF1V) to the spectra of the two MK standard stars HD 23643 (A3V) and HD 40136 (F1V). Details on the spectroscopic observations are given in section \ref{spectroscopic_observations}.}
    \label{fig_HD81076_Am}
\end{figure}

A comparison of the blue-violet spectrum (3800--4600 \AA) of HD 81076 to the spectra of the two MK standard stars HD 23643 (A3V) and HD 40136 (F1V) is provided in Figure \ref{fig_HD81076_Am}. The Ca II K line of HD 81076 is comparable in strength to that of the A3V MK standard. On the other hand, the metallic-line spectrum of HD 81076 is much more pronounced and resembles that of the F1V MK standard star HD 40136. The hydrogen-line spectrum is of an intermediate type, estimated here to about A8. We deduce that HD 81076 is a classical CP1 star and derive a spectral type of kA3hA8mF1V.

\begin{figure}
	\includegraphics[width=\columnwidth]{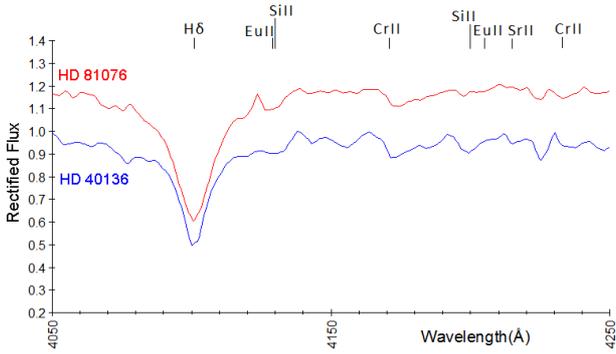}
    \caption{Detailed view of the regions around the H$\delta$ lines of HD 81076 (red) and HD 40136 (blue). The positions of some of the most prominent Eu II and Cr II features are indicated. See text for details.}
    \label{fig_HD81076_abundances}
\end{figure}

In order to further investigate possible overabundances in Eu and Cr, as suggested by \citet{1978mcts.book.....H}, we have compared the strength of some of the most prominent Eu II and Cr II features in HD 81076 and the F1V MK standard HD 40136 (cf. Fig. \ref{fig_HD81076_abundances}). Although the resolution of our spectra is not sufficient to resolve the investigated lines, substantial overabundances in Eu and Cr should result in significant blends around the indicated positions. These, however, are not observed, which makes us confident that at least no strong overabundances in Eu and Cr are present. This is further corroborated by the absence of a flux depression at around 5200\,\AA \,in the spectrum of HD 81076. This feature is a characteristic of the magnetic CP2 stars, which are known to exhibit strong Eu and Cr overabundances (cf. also section \ref{HD_98000_discussion}). It is interesting, though, that the Ca II K line of HD 81076 appears to have a rather peculiar profile (cf. Fig. \ref{fig_HD81076_Am}), which is a characteristic sometimes observed in CP2 stars. This feature, if real, will be investigated in more detail in the upcoming study, which will be based on high-resolution spectroscopy.

We have analysed photometric data from the SWASP project (9401 observations; time base of $\sim$717 days), the ASAS-3 archive (953 observations; time base of $\sim$3300 days), and our own observations (773 observations; time base of $\sim$11.3 days). A check of the corresponding sky region using the ALADIN interactive sky atlas indicates no star of sufficient brightness near our target that might affect the sky survey measurements.

Except for the first frequency, the observed amplitude of variation is very small. Thus, only the dominant frequency is traceable in ASAS-3 data. The same holds true for ROAD data, which -- after removing a signal near 3\,c/d from the dataset, which is obviously spurious and due to the intrinsic characteristics of the dataset -- yield the same frequency within the errors. Therefore, and because of the short time base of our own observations, our analysis has been primarily based on the SWASP dataset, which is of excellent quality and better time resolution.

\begin{figure}
	\includegraphics[width=\columnwidth]{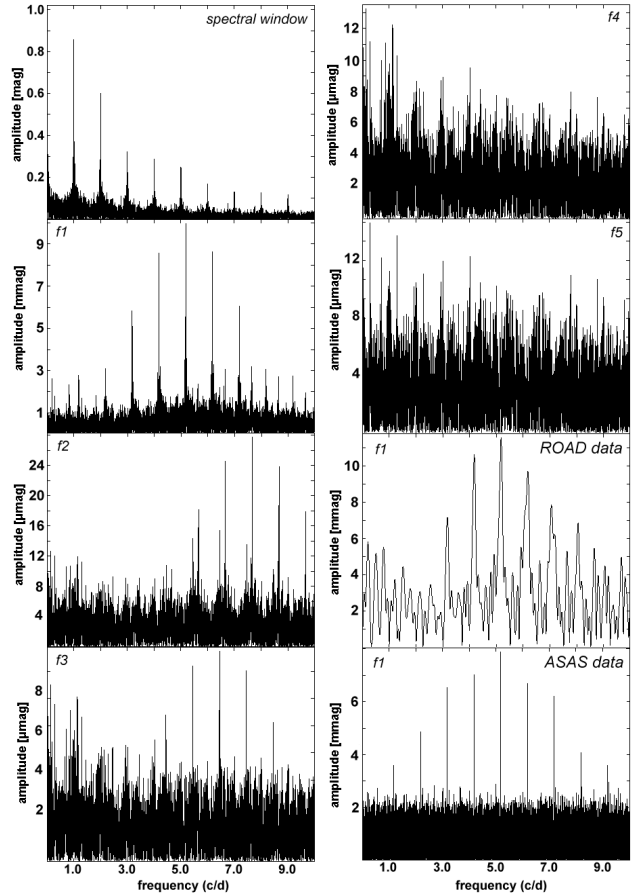}
    \caption{Period analysis of SWASP and ASAS data for HD 81076, illustrating the different steps of the frequency spectrum analysis. The top left panel shows the spectral window of SWASP data, which is dominated by daily aliases. The other left panels and the top two right panels show the frequency spectrum for unwhitened SWASP data and data that has been prewhitened with $f1$, $(f1+f2)$, $(f1+f2+f3)$, and $(f1+f2+f3+f4)$, respectively. The lower two right panels show the frequency spectra for unwhitened ROAD and ASAS data, respectively.}
    \label{fig_HD81076_pa}
\end{figure}

\begin{figure}
	\includegraphics[width=\columnwidth]{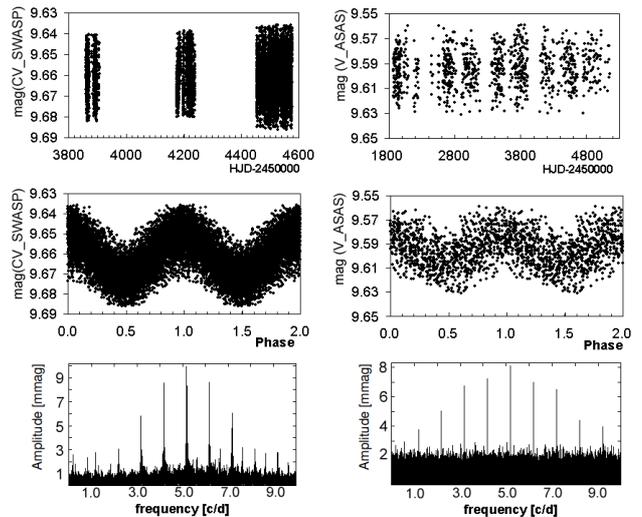}
    \caption{Light curves, phase plots and unwhitened frequency spectra of HD 81076, based on SWASP data (left panels) and ASAS-3 data (right panels).}
    \label{fig_HD81076_pp}
\end{figure}

SWASP data for HD 81076 are made up of three distinct parts separated by observational gaps of varying length (cf. Fig. \ref{fig_HD81076_pp}). The corresponding parts of the data have slightly different zero points; consequently, the light curve was detrended by shifting all parts of the data to the mean magnitude of the combined dataset. The results of our data analysis are given in Figure \ref{fig_HD81076_pa} and Table \ref{HD81076_table}. Data distribution and phase plots, folded with the derived best fitting frequencies, are illustrated in Fig. \ref{fig_HD81076_pp}.

\begin{table}
\begin{center}
\caption{HD 81076; frequencies and corresponding semi-amplitudes detected in the SWASP, ASAS-3, and ROAD datasets, as derived with PERIOD04.}
\label{HD81076_table}
\begin{adjustbox}{max width=0.5\textwidth}
\begin{tabular}{ccc}
\hline
\hline
ID & frequency & semi-amplitude \\
   & [c/d]     & [mag] \\       
\hline
$f1\textsubscript{SWASP}$	& 5.183410(7) & 0.0098 \\
$f2\textsubscript{SWASP}$	& 7.6564(2) & 0.0028 \\
$f3\textsubscript{SWASP}$	& 6.4500(4) & 0.0015 \\
$f4\textsubscript{SWASP}$	& 0.1317(5)$^{1}$ & 0.0014 \\
$f5\textsubscript{SWASP}$	& 0.2862(4)$^{1}$ & 0.0012 \\
\hline
$f1\textsubscript{ASAS}$	& 5.18340(1) & 0.0082 \\
\hline
$f1\textsubscript{ROAD}$	& 5.183(5) & 0.0116 \\
\hline
\multicolumn{3}{l}{$^{1}$ Likely spurious. See text for details.}
\end{tabular}
\end{adjustbox}
\end{center}
\end{table}

Five peaks were identified in the SWASP data. The first frequency corresponds within the errors to the first frequencies in the other datasets ($f1\textsubscript{SWASP}$ $\sim$ $f1\textsubscript{ASAS}$ $\sim$ $f1\textsubscript{ROAD}$), which is proof of its significance. However, the interpretation of the remaining peaks is not as straightforward. Before the background of the observed zero point differences mentioned above, we have checked the influence of seasonal effects on the Fourier analysis and the detection of possibly spurious frequencies in SWASP data. Furthermore, periodograms based on SWASP data are known to suffer from low-frequency noise (\citealt{2014MNRAS.439.2078H}; Barry Smalley, private communication); thus care has to be taken in the interpretation of the low frequencies $f4$ and $f5$.

In order to investigate these issues, SWASP data were split into three parts at the observational gaps (part 1: HJD 2453860--2453907, time span $\sim$47 d; part 2: HJD 2454175--2454239, time span $\sim$64 d; part 3: HJD 2454453--2454577, time span $\sim$124 d; cf. Fig. \ref{fig_HD81076_pp}), and each single observing season was searched for periodic signals. Interestingly, only $f1$ and $f2$ were recovered in each part of the dataset. The frequencies $f3$ to $f5$ were not detected in the first two parts of the SWASP data. The third part corresponds to the part with the longest consecutive coverage, which might go some way in explaining the observed discrepancy. Nevertheless, this casts doubt on the reality of these frequencies.

The frequency $f3$ is a clear detection in the third part of the data, and it is in obvious relationship to $f1$ and $f2$ (see discussion below). Therefore, we are inclined to interpret $f3$ as an intrinsic feature of the star's variability pattern. However, independent confirmation is needed.

The low frequencies $f4$ and $f5$ are in no obvious relationship to the higher frequencies and their combinations. We thus rule out the possibility that they originate in combination or aliasing issues. Furthermore, both frequencies are in the range of rotation frequencies encountered in A-type CP stars. However, in light of their non-detection in the first two parts of the SWASP dataset, their borderline low signal-to-noise ratio, and the low-frequency noise inherent to many SWASP datasets, their reality is open to question. In order to further investigate this issue, we have analysed SWASP data for some nearby (non-variable) stars (1SWASP J092308.93-453947.2, 1SWASP J092308.65-454018.0) and also find significant low-frequency noise in these datasets. Taking into account the evidence discussed above, we conclude that the low frequencies $f4$ and $f5$ are very likely spurious detections. We will therefore limit the following discussion to the first three frequencies detected in the SWASP dataset.

Some interesting relationships are apparent between the three highest peaks in SWASP data ($f1$, $f2$, and $f3$), which are almost equally spaced with $f2-f3$\,=\,1.2065 and $f3-f1$\,=\,1.2665. Since the spectral window of the SWASP data set (Fig. \ref{fig_HD81076_pa}, top panel) does not show any suspicious features near 1.20 or 1.27\,c/d, we consider this spacing as intrinsic to the variability pattern of the star. This closely-spaced triplet of frequencies might be explained by rotational splitting of non-radial oscillation modes \citep[e.g.][]{2010ApJ...721.1900D}.

In any case, pulsation as the underlying mechanism for the three highest frequencies, which are situated at the low frequency end of the expected values for DSCT stars, seems secure. It is highly unlikely that these frequencies are caused by rotational modulation as the corresponding rotational period would be unrealistically short for a late A- / early F-type main sequence star. Considering the observed time scales of variability and the star's position in the $M_{\rm Bol}$ versus $\log T_\mathrm{eff}$ diagram (cf. Figure \ref{hrd}), we conclude that the CP1 star HD 81076 is a DSCT variable.

\subsubsection{HD 98000} \label{HD_98000_discussion}

HD 98000 was classified as spectral type B9 in the Henry Draper Catalogue \citep{1919AnHar..94....1C} and later reclassified as ApSi by \citet{1973AJ.....78..687B} and \citet{1978mcts.book.....H}. On the basis of this information, the star was included in the RM09 catalogue with a spectral type of B9pSi. It has never been subjected to a variability study before.

\begin{figure}
	\includegraphics[width=\columnwidth]{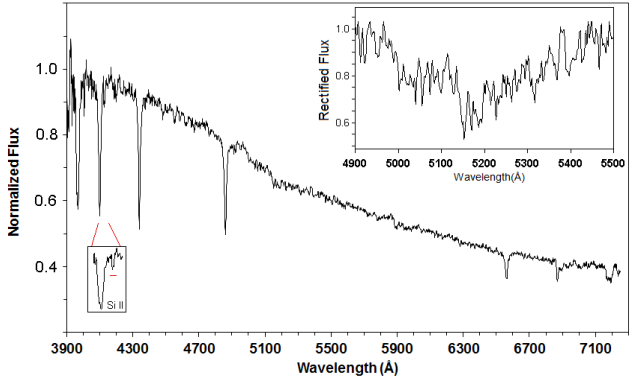}
    \caption{The Mirranook Observatory classification resolution spectrum of HD 98000 (3900\,\AA\,<\,$\lambda$\,<\,7300\,\AA). The insets illustrate the flux depression at around 5200 \AA \,(upper right), which is characteristic of magnetic CP2 stars, and the strong blend likely due to Si II $\lambda \lambda$4128-30\,\AA. Details on the spectroscopic observations are given in section \ref{spectroscopic_observations}.}
    \label{fig_HD98000_spectrum}
\end{figure}

The classification resolution spectrum of HD 98000 is shown in Fig. \ref{fig_HD98000_spectrum}. It confirms the nature of HD 98000 as a classical CP2 star of the silicon subgroup. Note in particular the flux depression at around 5200 \AA, which is characteristic of the magnetic CP2 stars \citep[e.g.][]{2003MNRAS.341..849K}. On the basis of our spectrum, we deduce a spectral type of B9pSi for HD 98000.

We have analysed photometric data from the ASAS-3 archive (703 observations; time base of $\sim$3300 days) and our own observations (2099 observations; time base of $\sim$21.2 days). An investigation of the corresponding sky region using the ALADIN interactive sky atlas indicates no star of sufficient brightness near our target which might affect the ASAS measurements. Because of the low amplitude of the observed photometric variability, ROAD measurements of HD 98000 were acquired through a clear filter in order to maximise throughput.

\begin{figure}
	\includegraphics[width=\columnwidth]{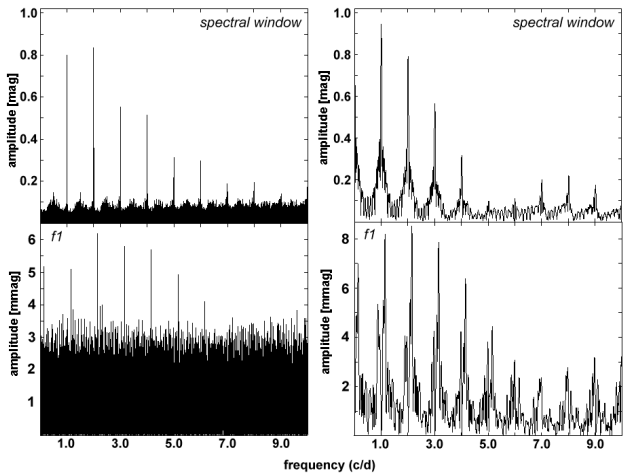}
    \caption{Period analysis of ASAS (left panels) and ROAD data (right panels) for HD 98000. The top panels show the spectral windows dominated by daily aliases. The bottom panels show the frequency spectra for unwhitened data.}
    \label{fig_HD98000_pa}
\end{figure}

\begin{figure}
	\includegraphics[width=\columnwidth]{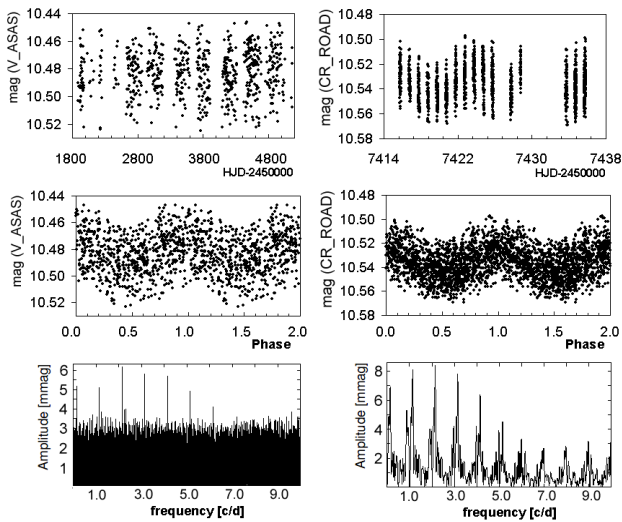}
    \caption{Light curves, phase plots and unwhitened frequency spectra of HD 98000, based on ASAS-3 data (left panels) and ROAD data (right panels).}
    \label{fig_HD98000_pp}
\end{figure}

Only one significant peak of low amplitude was identified in ROAD and ASAS-3 data; within the errors, the derived frequency is the same in both datasets, which is proof of its reality. No other frequencies were identified; we estimate the threshold of detection to be $\sim$0.004\,mag and $\sim$0.003\,mag for ASAS and ROAD data, respectively. The results of our period analysis are given in Figure \ref{fig_HD98000_pa} and Table \ref{HD98000_table}. Data distribution and phase plots, folded with the derived best fitting frequencies, are illustrated in Fig. \ref{fig_HD98000_pp}.

\begin{table}
\begin{center}
\caption{HD 98000; frequency and corresponding semi-amplitude detected in the ROAD and ASAS-3 datasets, as derived with PERIOD04.}
\label{HD98000_table}
\begin{adjustbox}{max width=0.5\textwidth}
\begin{tabular}{ccc}
\hline
\hline
ID & frequency & semi-amplitude \\
   & [c/d]     & [mag] \\       
\hline
$f1\textsubscript{ASAS}$	& 2.14751(2) & 0.0061 \\
\hline
$f1\textsubscript{ROAD}$	& 2.145(1) & 0.0116 \\
\hline
\end{tabular}
\end{adjustbox}
\end{center}
\end{table}
	
Without a doubt, HD 98000 is a classical CP2 star of the silicon subgroup. The only proven form of pulsational variability among this type of CP stars is observed in the so-called rapidly oscillating Ap (roAp) stars \citep{1982MNRAS.200..807K} which exhibit photometric variability in the period range of $\sim$5-20\,min (high-overtone, low-degree, and non-radial pulsation modes). This is very different from what has been observed for HD 98000. \citet{2016A&A...588A..71E} have shown that other kinds of pulsational variability might exist in magnetic CP stars; however, the object investigated by the aforementioned authors has evolved significantly away from the Terminal Age Main Sequence and is thus no classical CP2 star and presumably in a very different evolutionary state than HD 98000.

Generally, photometric variability in CP2 stars is considered to be caused by rotational modulation due to the redistribution of flux in the surface abundance spots \citep[e.g.][]{2013A&A...556A..18K}. Hence the observed photometric period is the rotational period of the star; photometrically variable CP2 stars are traditionally referred to as ACV variables. ACV variables are in almost all cases monoperiodic, which -- in the accuracy limit of the available data -- is also true for HD 98000.

From the available data, we propose rotational modulation as the underlying cause of the observed photometric variability in the monoperiodic, classical CP2 star HD 98000. \citet{2010A&A...511L...7M} provide a list of the fastest rotating CP2 stars discovered to this date, whose rotational periods cluster around a value of $P \approx$ 0.52\,d (HR 7355, $P$ = 0.52144\,d; HD 164429, $P$ = 0.51899\,d; CU Vir, $P$ = 0.52070\,d; HD 92385, $P$ = 0.54909\,d). With a period of $P \approx$ 0.466\,days, HD 98000 would be the ACV variable with the shortest period hitherto observed.

However, as ACV variables are prone to exhibit double-wave variations \citep{1980A&A....89..230M} and the observed amplitude of variability is very small for HD 98000, a twice longer rotation period cannot be definitely excluded. Furthermore, due to the strong aliasing inherent to both datasets, we cannot exclude the possibility that our derived period is an alias of the true value. However, as the proposed frequency has been unambiguously identified in both datasets, and the data folded on $(f/2)$ do not show a convincing phase plot, current evidence favours the period solution proposed above.

If our assumption concerning the origin of the photometric variability and the derived period is proven to be true, HD 98000 is the ACV variable with the shortest period hitherto observed and hence a very interesting object that might help to investigate the influence of rotational mixing on chemical peculiarities and define an upper limit for rotational velocities, beyond which no chemical peculiarities due to the interplay of radiative diffusion and gravitational settling are expected. We therefore strongly encourage further photometric and phase-resolved spectroscopic observations of HD 98000 in order to unravel the underlying mechanism of variability.

\section{Conclusions} \label{conclusions}
We have carried out an investigation of four photometrically variable chemically peculiar stars. New spectroscopic observations were acquired which confirm the peculiar composition of all targets. HD 77013 and HD 81076 are classical CP1 (Am) stars; HD 67983 is a marginal CP1 (Am:) star, and HD 98000 is a CP2 (Ap) star showing enhanced lines of silicon. We have derived astrophysical parameters and investigated our sample stars in the $M_{\rm Bol}$ versus $\log T_\mathrm{eff}$ diagram, from which we derive information on evolutionary status. We have discussed each object and presented detailed period analyses that have been based on publicly available data from the ASAS-3 and SuperWASP archives as well as our own photometric observations.

Taking into account all available data, we propose pulsations as the underlying mechanism of the variability observed in HD 67983, HD 77013 and HD 81076. HD 67983 and HD 77013 exhibit multiperiodic variability in the $\gamma$ Doradus frequency realm; HD 81076 was identified as a $\delta$ Scuti star. We intend to obtain high-resolution spectroscopic observations of our target stars, which will be presented in a forthcoming paper.

The CP2 star HD 98000 exhibits monoperiodic variability with a frequency of $f \approx$ 2.148 c/d ($P \approx$ 0.466 d), which we interpret as the rotational period. If our assumption is verified, HD 98000 is the ACV variable with the shortest period hitherto observed and thus a very interesting object that might help to investigate the influence of rotational mixing on chemical peculiarities. We therefore strongly encourage phase-resolved observations of HD 98000 in order to unravel the underlying mechanism of variability.

\section*{Acknowledgements}

We thank Barry Smalley and Daniel Holdsworth for helpful discussions on the interpretation of SuperWASP data, and the referee for helpful suggestions and advice that helped to improve the paper. This publication is based (in part) on spectroscopic data obtained through the collaborative Southern Astro Spectroscopy Email Ring (SASER) group. This research has made use of the SIMBAD and VizieR databases operated at the Centre de Donn\'ees Astronomiques (Strasbourg) in France. Furthermore, data products from the Two Micron All Sky Survey were employed, which is a joint project of the University of Massachusetts and the Infrared Processing and Analysis Center/California Institute of Technology, funded by the National Aeronautics and Space Administration and the National Science Foundation. This work has also made use of data from the European Space Agency (ESA) mission {\it Gaia} (\url{http://www.cosmos.esa.int/gaia}), processed by the {\it Gaia} Data Processing and Analysis Consortium (DPAC, \url{http://www.cosmos.esa.int/web/gaia/dpac/consortium}). Funding for the DPAC has been provided by national institutions, in particular the institutions participating in the {\it Gaia} Multilateral Agreement.


\bsp	
\label{lastpage}
\end{document}